\documentclass{article}

\usepackage[numbers]{natbib}         
\usepackage[colorlinks]{hyperref}    
\usepackage[english]{babel} 
\usepackage{amssymb}
\usepackage{amsmath}
\usepackage{txfonts}
\usepackage{mathdots}
\usepackage[classicReIm]{kpfonts}
\usepackage[dvips]{graphicx} 
\usepackage[a4paper, portrait, margin=1in]{geometry}
\usepackage{tabularx}
\usepackage{longtable}
\usepackage{multirow}
\usepackage{booktabs}
\usepackage[labelsep=period]{caption}
\usepackage{makecell}
\begin{document}
	\setlength{\parindent}{0pt}
	\setlength{\parskip}{1ex}
	
	\textbf{\Large Intensity Non-uniformity Correction in MR Imaging Using Residual Cycle Generative Adversarial Network}
	
	\bigbreak

	Xianjin Dai$^1$, Yang Lei$^1$, Yingzi Liu$^1$, Tonghe Wang$^1$, Lei Ren$^2$, Walter J. Curran$^1$, Pretesh Patel$^1$, Tian Liu$^1$ and Xiaofeng Yang$^1$*
	
	$^1$Department of Radiation Oncology and Winship Cancer Institute, Emory University, Atlanta, GA 30322
	
	$^2$Department of Radiation Oncology, Duke University, Durham, NC 27708

	\bigbreak
	\bigbreak
	\bigbreak

	\textbf{*Corresponding author: }
	
	Xiaofeng Yang, PhD
	
	Department of Radiation Oncology
	
	Emory University School of Medicine
	
	1365 Clifton Road NE
	
	Atlanta, GA 30322
	
	E-mail: xiaofeng.yang@emory.edu

	\bigbreak
	\bigbreak
	\bigbreak
	\bigbreak
	\bigbreak
	\bigbreak

	\textbf{Abstract}

	\textbf{Purpose:} Correcting or reducing the effects of voxel intensity non-uniformity (INU) within a given tissue type is a crucial issue for quantitative MRI image analysis in daily clinical practice. Although having no severe impact on visual diagnosis, the INU artifact can highly degrade the performance of automatic quantitative analysis such as segmentation, registration, feature extraction and radiomics. In this study, we present a deep learning-based approach for MRI image INU correction.
	
	\textbf{Method:} An advanced deep learning based INU correction algorithm called residual cycle generative adversarial network (res-cycle GAN), which integrates the residual block concept into a cycle-consistent GAN (cycle-GAN), was developed in this work. In cycle-GAN, an inverse transformation was implemented between the INU uncorrected and corrected MRI images to constrain the model through forcing the calculation of both an INU corrected MRI and a synthetic corrected MRI. A fully convolution neural network integrating residual blocks was applied in the generator of cycle-GAN to enhance end-to-end raw MRI to INU corrected MRI transformation. A cohort of 30 abdominal patients with T1-weighted MR INU images and their corrections with a clinically established and commonly used method, namely, N4ITK were used as a pair to evaluate the proposed res-cycle GAN based INU correction algorithm. Quantitatively comparisons of normalized mean absolute error (NMAE), peak signal-to-noise ratio (PSNR), normalized cross-correlation (NCC) indices, and spatial non-uniformity (SNU) were made among the proposed method and other approaches. 
	
	\textbf{Result:} Our res-cycle GAN based method achieved higher accuracy and better tissue uniformity compared to the other algorithms, in terms of NMAE, PSNR, NCC outcomes. Overall, improvements in NMAE of 39\%, PSNR of 17\%, NCC of 1.5\%, and SNU of 47\% were found. Moreover, once the model is well trained, our approach can automatically generate the corrected MR images in a few minutes, eliminating the need for manual setting of parameters. 
	
	\textbf{Conclusion:} In this study, a deep learning based automatic INU correction method in MRI, namely, res-cycle GAN has been investigated. The results show that learning based methods can achieve promising accuracy, while highly speeding up the correction through avoiding the unintuitive parameter tuning process in N4ITK correction.
	
	\bigbreak
	\bigbreak
	
	\textbf{keywords:} Magnetic Resonance Imaging (MRI), bias field, intensity non-uniformity, deep learning, generative adversarial network (GAN).

	\noindent 
	\section{ INTRODUCTION}
	
	Magnetic resonance imaging (MRI) is an established non-invasive three-dimensional (3D) imaging technique, which is widely used in diagnosis and therapy due to its capability in providing meaningful anatomical information.\cite{RN3818, RN3821, RN3825, RN3827, RN3837}. The applications of MRI in radiation therapy treatment planning have been increased in the past decade because of its superb soft-tissue contrast over the conventionally used X-ray computed tomography (CT).\cite{RN3797, RN3801, RN3829} In addition, with its capability of characterizing the tumor phenotype by using advanced image analysis techniques such as radiomics,\cite{RN3834} MRI plays an important role in personalized precision radiation therapy. In these applications, it is very important to extract the information essential for diagnosis and therapy precisely and accurately. As an urgent need to handle the rapidly increasing volumes of image data from longitudinal studies, clinical trials, and clinical practices, significant advances have taken place in the field of automated image analysis algorithms such as segmentation, classification, registration and image synthesis.\cite{RN3793, RN3805, RN2641, RN1695, RN1697, RN1676, RN1694, RN3833, RN1703, RN1788, RN1427} A robust, reliable, and inexpensive automated image analysis pipeline is highly desired. To achieve this goal, in MRI, offering artifact-free images is the fundamental first step. However, it is difficult to avoid artifacts during the MRI image acquisition. Therefore, an image preprocessing step is necessary to eliminate or mitigate the artifacts. 
	
	Typical MR images have artifacts from different sources such as patient motion- and machine/hardware-induced. One of the most common artifacts is intensity non-uniformity (INU) or bias field, which refers to the slow, nonanatomic intensity variations of the same tissue over the image.\cite{RN3796, RN3798, RN3803, RN3812, RN3816, RN3835, RN3839} It is usually caused by static field inhomogeneity, radiofrequency coil non-uniformity, gradient-driven eddy currents, inhomogeneous reception sensitivity profile, and overall subject’s anatomy both inside and outside the field of view. While low level intensity non-uniformity artifact (intensity variations less than 30\%) might have little impact on visual diagnosis, the performance of automatic image analysis approaches can be significantly degraded by clinically acceptable levels of intensity non-uniformity, due to the fact that homogeneity of intensity within each class is the primary assumption in these automatic image analysis methods. Early studies on INU correction date back to 1986.\cite{RN3808, RN3822} Since then, extensive studies have been carried out and lots of approaches have been proposed on INU correction.\cite{RN3796, RN3820, RN3832, RN3835} In general, two types of methods have been widely investigated, (a) prospective calibration and (b) retrospective correction methods. The prospective calibration methods are generally intended to model INU as hardware-related factors and correct INU through compensation by acquiring supplementary images of uniform phantoms,\cite{RN3795} integrating information from different coils,\cite{RN1587} and designing dedicated imaging sequences.\cite{RN3799} Those prospective methods can certainly eliminate hardware-related inhomogeneities, nevertheless, they are hardly solving subject-related nonuniformity. To tackle this challenge, retrospective correction methods which rely on image features to remove spatial nonuniformity have been proposed and widely used nowadays.\cite{RN3794, RN3817, RN3819, RN3823, RN3831} Theoretically, retrospective corrective methods account for both hardware and subject induced INU components.\cite{RN3804} Among those retrospective correction methods, N4ITK is currently the most well-known, efficient, and the most commonly used method for INU correction in MR imaging.\cite{RN3820, RN3832} However, in practice, it requires unintuitive patient-specific parameter tuning and is usually time expensive. Thus, it is highly desired to develop an efficient INU correction algorithm which is patient-specific parameter independent. As technical advances in deep learning, a few studies have been carried out to develop highly automated INU correction approaches using deep learning.\cite{RN3830, RN3836} While inspiring, conventional networks were used and the performance of INU correction was relatively limited. In this study, an advanced deep learning algorithm, namely, residual cycle generative adversarial network (res-cycle GAN) \cite{RN3807, RN3814} has been investigated to further improve the performance of INU correction in MR imaging.

	\noindent 
	\section{Methods and Materials}
	\noindent 
	\subsection{Overview}
	
	A res-cycle GAN integrating cycle-consistent GAN and residual learning is proposed to capture the relationship between uncorrected raw MRI and INU corrected MRI images. Cycle-consistent GAN was originally investigated for image-to-image translation.\cite{RN3800, RN3813, RN3815, RN3838} By learning a mapping G:X→Y such that the distribution of images from G(X) is indistinguishable from the distribution Y using an adversarial loss, cycle-consistent GAN, has archived attractive performance for image-to-image translation when the paired images are absent. In this work, the same learning strategy was adapted. But, paired raw MRI and INU corrected MRI data were used for training the network. Moreover, unlike conventional cycle-consistent GAN cases, the source images (uncorrected raw MRI) and target images (INU corrected MRI) are largely similar in our study. Therefore, learning the residual image which is the difference between the source and target images, rather than the entire images, can enhance the convergence thus improve the efficiency of training.\cite{RN3809, RN3810} As shown in Fig. 1, the proposed res-cycle GAN based MRI INU correction method consists of a training stage and a correction stage. In the training stage, the N4ITK corrected MRI was used as the learning target of the raw MRI image since N4ITK is a clinically established and the most frequently used method for MRI INU correction.  
	Each component of the network (as shown in Fig. 1) is outlined in further detail in the following sections.

	\begin{figure}
		\centering
		\noindent \includegraphics*[width=6.50in, height=4.20in, keepaspectratio=true]{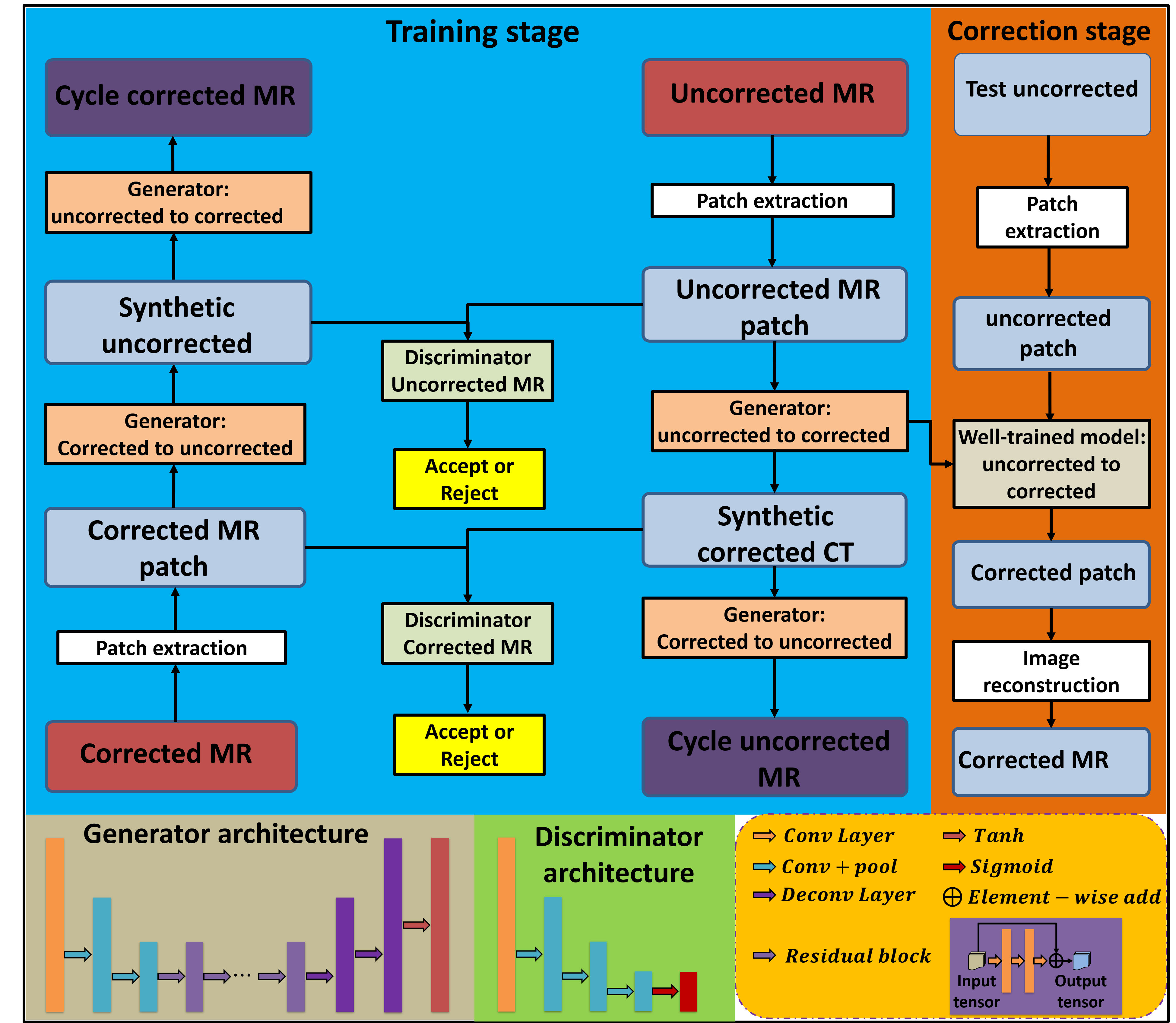}
		
		\noindent Fig. 1. The schematic flow diagram of the res-cycle GAN for MRI INU correction.
	\end{figure}

	\noindent 
	\subsection{Residual cycle GAN}
	
	Conventional generative adversarial nets (GANs) primarily rely on the idea of an adversarial loss, in which, a generative model and a discriminative model work against each other \cite{RN3806}. The generative model works as a counterfeiter trying to produce fake currency, while, the discriminative model detects the counterfeit currency. Competition between the generator and discriminator drives both teams to improve their performance until the counterfeits are indistinguishable from the genuine article. GANs have been widely investigated for image-to-image translation, learning the mapping such that the translated images cannot be distinguished from images in the target domain. Cycle-consistent GAN (cycle-GAN) was originally proposed for image-to-image translation in case that pair source and target images are absent. In this work, a similar idea of cycle-GAN was adapted. Given the paired datasets of raw MRI which has INU artifacts and INU corrected MRI, an initial mapping is learned to generate an INU corrected MRI-like image from an uncorrected raw MRI, called synthetic corrected MRI, which is largely similar to the true INU corrected MRI to fool the discriminative model. Conversely, the discriminator is trained to differentiate INU corrected MRI from synthetic corrected MRI. As the discriminators and the generators, these networks are pitted against each other, the capabilities of each improve, leading to more accurate synthetic corrected MRI generation. The entire network is optimized sequentially in a zero-sum framework. Compared to conventional GANs, inverse transformations include translating a raw MRI with INU artifact to a synthetic corrected MRI, and a corrected MRI to a synthetic uncorrected MRI, are introduced to constrain the model thus increase accuracy in output images.
	Since the raw MRI images with INU artifact and INU corrected MRI images are largely similar, the strategy of residual learning is adopted and integrated in the cycle-consistent GAN architecture. Residual learning is originally investigated for improving the convergent efficiency during training and the performance of deep neural networks when increasing the network depth.\cite{RN3809} Particularly, residual learning has achieved promising results in tasks where input and output images are largely similar,\cite{RN3810} much like the relationship between raw MRI with INU artifacts and INU corrected MRI images. As shown in Figure 1, in each generator, the feature map is firstly reduced in size by two down-sampled convolution layers, followed by nine short-term residual blocks, and then two deconvolution layers and a tanh layer for obtaining the output map. Each residual block consists of a residual connection, two hidden convolution layers, and an element-wise summation operator. In our res-cycle GAN architecture, each discriminator consists of one convolutional layer, three convolutional with pooling layers, and a sigmoid function.

	\noindent 
	\subsection{Compound loss function}
	
	A two-part loss function including an adversarial loss and a cycle consistency loss was used to optimize learnable parameters in the original cycle-consistent GAN.\cite{RN3838} The adversarial loss function, which relies on the output of the discriminators, applies to both the INU uncorrected MRI to the corrected MRI generator $G_{u-c}$) and the INU corrected MRI to uncorrected MRI generator $G_{c-u}$, but here we present only formulation for $G_{u-c}$ for clarity. The adversarial loss function is defined as follow:
	\begin{equation} 
	L_{adv}\ (G_{u-c},D_{c},I_{u},I_{c})\ =SCE[D_{c}\ (G_{u-c}\ (I_{u})),1]
	\end{equation} 
	where $I_{u}$ is the INU uncorrected MRI image and $G_{u-c}(I_{u})$ is the output of the uncorrected MRI to corrected MRI generator. $D_{c}$ is the corrected MRI discriminator which is designed to return a binary value indicating whether a pixel region is real (corrected MRI) or fake (synthetic corrected MRI), thus it measures the number of incorrectly generated pixels in the synthetic corrected MRI image. The function $SCE\left(\bullet,1\right)$ is the sigmoid cross entropy between the discriminator map of the generated synthetic corrected MRI and a unit mask.  
	
	The cycle consistency loss function consists of a compound loss function, which forces the transformation from uncorrected MRI to INU corrected MRI to be close to a one-to-one mapping. In this work, the cycle consistency loss function is applied for the generators of uncorrected MRI to cycle uncorrected MRI and corrected MRI to cycle corrected MRI, respectively. Besides these constraints, the distance between synthetic corrected MRI (fake) and INU corrected MRI (real), and the distance between synthetic uncorrected MRI (fake) and uncorrected MRI (real) are minimized by what we called the synthetic consistency loss function, respectively. The synthetic consistency loss function can directly enforce the real image and its fake image to have the same intensity distribution. The first component of the combined cycle- and synthetic-consistency loss function is the mean absolute loss (MAL): 
	\begin{equation} 
	MAL(G_{u-c},G_{c-u})=\frac{1}{n(I_u)}\left[\begin{matrix}\lambda_{loss}^{cycle}\left[\begin{matrix}||G_{c-u}[G_{u-c}(I_u)]-I_u||_1\\+||G_{u-c}[G_{c-u}(I_c)]-I_c||_1\end{matrix}\right]\\+\lambda_{loss}^{syn} \left[\begin{matrix}||G_{c-u}(I_c)-I_u||_1\\+||G_{u-c}(I_u)-I_c||_1\end{matrix}\right]\\\end{matrix}\right]                    
	\end{equation} 
	where $n(\bullet)$ is the total number of pixels in the image, $\lambda_{loss}^{cycle}$ and $\lambda_{loss}^{syn}$ are parameters which control the cycle consistency and synthetic consistency, respectively. The symbol $||\bullet||_1$  signifies the L1-norm for a vector x, i.e. mean absolute error.
	The second component of the combined cycle- and synthetic-consistency loss function is the gradient-magnitude distance (GMD), which is defined as:
	\begin{equation} 
	\begin{split}
	&GMD(Z,Y)=\\ &\sum_{i,j,k}\left\{\begin{matrix}\left(\left|Z_{i,j,k}-Z_{i-1,j,k}\right|-\left|Y_{i,j,k}-Y_{i-1,j,k}\right|\right)^2+\left(\left|Z_{i,j,k}-Z_{i,j-1,k}\right|-\left|Y_{i,j,k}-Y_{i,j-1,k}\right|\right)^2\ \\+\ \left(\left|Z_{i,j,k}-Z_{i,j,k-1}\right|-\left|Y_{i,j,k}-Y_{i,j,k-1}\right|\right)^2\\\end{matrix}\right\}  
	\end{split}                       
	\end{equation}
	
	where Z and Y are any two images, and i, j, and k represent pixels in x, y, and z. 
	Besides, a gradient magnitude loss (GML), which is a function of the generator networks, is defined as:
	
	\begin{equation}
		GML\left(G_{u-c},G_{c-u}\right)=\lambda_{dist}^{cycle}\left[\begin{matrix}GMD(G_{c-u}[G_{u-c}(I_u)],I_u)\\+GMD(G_{u-c}[G_{c-u}(I_c)],I_c)\\\end{matrix}\right]+       \lambda_{dist}^{syn}\left[\begin{matrix}GMD(G_{c-u}(I_c),I_u)\\+GMD(G_{u-c}(I_u),I_c)\\\end{matrix}\right]                
	\end{equation}
	The combined cycle- and synthetic-consistency loss function is then optimized by:
	\begin{equation}
	L_{cyc,\ syn}(G_{u-c},G_{c-u})\ =MAL(G_{u-c},G_{c-u})+ GML(G_{u-c},G_{c-u})               
	\end{equation}
	The global generator loss function can then be written as:
	\begin{equation}
	L_{GEN}\left({G_{c-u},G}_{u-c}\right)=\underset{G_{c-u},G_{u-c}}{arg\min}\left\{\lambda_{adv}\left[\begin{matrix}L_{adv}(G_{u-c},D_c,I_u,I_c)\\+L_{adv}(G_{c-u},D_u,I_c,I_u)\\\end{matrix}\right]+L_{cyc,\ syn}(G_{u-c},G_{c-u})\right\}               
	\end{equation}
	where${\lambda}_{adv}$ is a regularization parameter that controls the weights of the adversarial loss. The discriminators are optimized in tandem with the generators by:
	\begin{equation}
	L_{DIS}\left(D_c,D_u\right)=\underset{D_c,D_u}{arg\min}\left\{\begin{matrix}SCE[D_c(G_{u-c}(I_u)),0]+SCE[D_c(I_c),1]\\+\\SCE[D_u(G_{c-u}(I_c)),0]+SCE[D_u(I_u),1]\\\end{matrix}\right\}               
	\end{equation}
	
	In our res-cycle GAN framework, the network is trained to generate images and differentiate between synthetic (fake) and real images simultaneously by learning the forward and inverse relationships between the source and target images. Given two images (uncorrected and corrected MRI) with similar underlying structures, the res-cycle GAN is designed to learn both intensity and textural mappings from a source distribution (uncorrected MRI) to a target distribution (corrected MRI).
	
	\noindent 
	\subsection{Image data acquisition and preprocessing}
	
	A retrospective study was conducted on a cohort of 30 abdominal patients with T1-weighted MRI images acquired on a GE Signa HDxt 1.5T MRI scanner. The sequence parameters were: TE ranging from 2.2 to 4.4 ms, TR ranging from 175 to 200 ms, patient position being FFS and flip angle being 80 degree. FOV was 480 mm × 480 mm and the acquisition length in axial direction was from 117 mm to 300 mm. The voxel size is 1.875 mm × 1.875 mm × 3.0 mm. Geometrical correction was performed using the built-in software package at the scanner. Respiratory belt was used to monitor breathing for the breath-hold scans. The scanning time for the T1 imaging was 20-26 seconds. After the MRI images were acquired, an open source software platform namely, 3D Slicer,\cite{RN3802, RN1692, RN3826} which was developed for medical image informatics, image processing, and three-dimensional visualization, was used for INU correction based on its integrated N4ITK algorithm. The corrected images were then used as the ground truth, and paired with their corresponding raw images for deep neural network training process.

	\noindent 
	\subsection{Implementation and performance evaluation}
	
	Since it’s hard to get ideal INU-free MRI images, a clinical established and commonly used INU correction algorithm N4ITK was adopted in this work for obtaining corrected MRI as the ground truth for training the res-cycle GAN. The uncorrected raw MRI and N4ITK corrected MRI images are firstly fed into the network in 64x64x64 patches without image thresholding. The res-cycle GAN was implemented based on the widely used deep learning framework Tensorflow.\cite{RN3792} The hyperparameter values for Eqs. (2) and (7) are listed as follows:$\lambda_{adv}=1$,$\lambda_{loss}^{cycle}=10$,$\lambda_{loss}^{syn}=1$.Typically, the value of $\lambda_{loss}^{cycle}$ should be larger than $\lambda_{loss}^{syn}$, because the two transformations incorporated in the architecture makes the generation of an accurate cycle image from a real image rather difficult. The learning rate for Adam optimizer in our algorithm was set to 2e-4, and the model is trained and tested on Ubuntu 18.04 environment with python 3.7. A batch size of 20 was used to fit the 32 GB memory of the NVIDIA Tesla V100 GPU. The network was trained for 70000 iterations and it took approximately 5.8 hours. Once the network was trained, the MRI correction of a new arrival patient’s data can be performed within 1 minutes, depending on image size. A 3D image can be formed by simply combing the patch output of the network. In the edge of patches that overlap upon output, the pixel values in the same position are averaged. 
	
	In this study, a leave-one-out cross validation approach was used for evaluating the proposed algorithm. Particularly, for each experiment, one patient’s images were used as test data, the rest patients’ images were used as training data. The model was trained on training data and test on test data. The experiment repeated 30 times to let each patient’s images used as test data exactly once. In addition to assessing the performance of our proposed method, we compare to both clinically established and commonly used N4ITK INU correction method and other INU correction methods based on two popular deep learning architectures, U-net\cite{RN3828} and conventional GAN.\cite{RN3806} For quantitative comparisons, normalized mean absolute error (NMAE), peak signal-to-noise ratio (PSNR), and normalized cross correlation (NCC) are calculated. NMAE is the normalized magnitude of the difference between the ground truth (INU corrected MRI) and the evaluated image, which can be formed as:
	
	\begin{equation}
	NMAE=\frac{1}{n_xn_yn_z}\sum_{i,j,k}^{n_x{n_yn}_z}\frac{|f\left(i,j,k\right)-t\left(i,j,k\right)|}{\underset{i,j,k}\max{i,j,k}{f\left(i,j,k\right)}}
	\end{equation}
	where $f\left(i,j,k\right)$ is the pixel value from ground truth, $t\left(i,j,k\right)$ is the value of pixel $\left(i,j,k\right)$ in the target image, and $n_x{n_yn}_z$ is the total number of pixels. 
	PSNR is calculated by:
	\begin{equation}
	PSNR=10\ \times\log_{10}{\left(\frac{MAX^2}{MSE}\right)}
	\end{equation}
	where MAX is the maximum signal intensity, and MSE is the mean-squared error of the image. 
	The NCC is a measure of the similarity of structures between two images. It is commonly used in image analysis and pattern recognition and is defined as \cite{RN36, RN37}:
	\begin{equation}
	NCC=\frac{1}{n_x{n_yn}_z}\sum_{i,j,k}^{n_xn_yn_z}{\frac{1}{\sigma_f\sigma_t}\left[f\left(i,j,k\right)\bullet t\left(i,j,k\right)\right]}
	\end{equation}
	where $\sigma_f$ is the standard deviation of ground truth, and $\sigma_t$ is the standard deviation of the target image. In addition, for quantitative assessment of intensity uniformity, a spatial non-uniformity (SNU) is defined and calculated. SNU is a measurement of the difference in image intensity within the region of interest (ROI), which is defined as:
	\begin{equation}
	SNU=\frac{I_{max}-I_{min}}{I_{mean}}
	\end{equation}
	where $I_{max}$, $I_{min}$, and $I_{mean}$ are the maximum, minimum, and mean intensity value in the ROI, respectively. To calculate SNU, first, the ROIs (as an example shown in Fig. 2) were selected. And, the maximum, minimum, and mean values were then calculated.
	\begin{figure}
		\centering
		\noindent \includegraphics*[width=6.50in, height=4.20in, keepaspectratio=true]{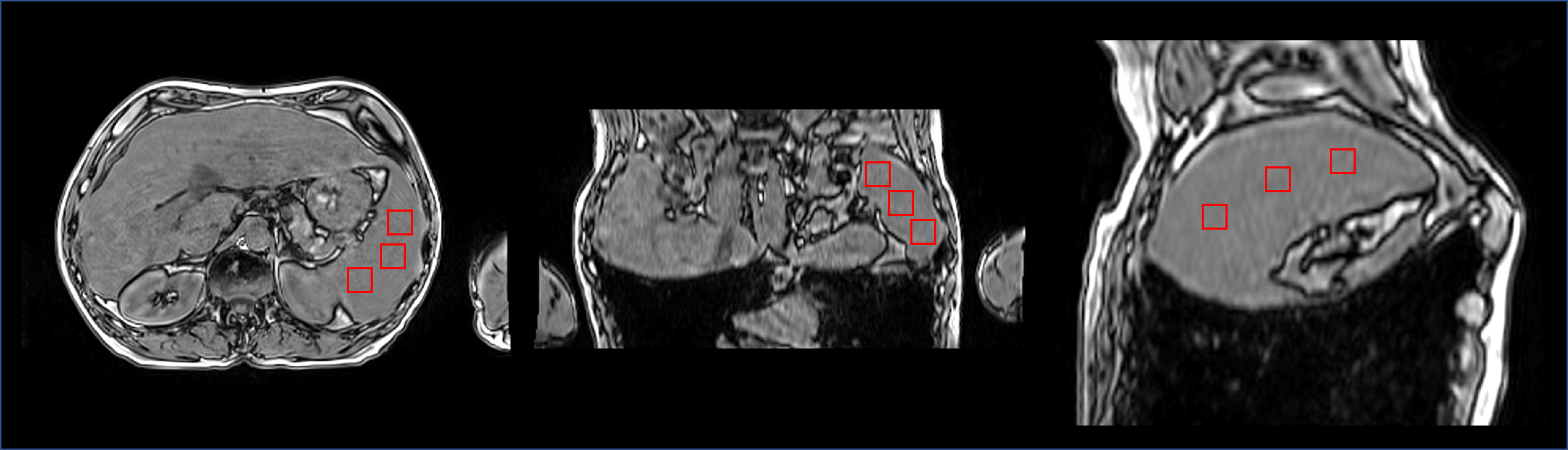}
		
		\noindent Fig.2 Selected ROIs for measuring spatial non-uniformity (SNU).
	\end{figure}
	All comparison metrics were calculated for each patient during evaluation. The entire dataset was used for quantitative analyses. Paired two-tailed t-tests were performed for comparing the outcomes between numerical results groups calculated from all patient data to demonstrate the statistical significance of quantitative improvement by our proposed algorithm. 
	\noindent 
	\section{Results}
	\subsection{Correction performance}
	
	Fig. 3 shows the results of INU correction on a patient test case. The top row shows the uncorrected, N4ITK corrected, and proposed algorithm corrected images, respectively. In this work, N4ITK corrected images were used as “ground truth” for training the networks, therefore, the differential images between the proposed algorithm and N4ITK correction were calculated, which is shown in right hand side of the top row in Fig. 3. By qualitatively comparisons of the image corrected by the proposed method to the uncorrected image, and N4ITK corrected image to uncorrected image, similar INU correction performance is seen between the proposed method and N4ITK. To further quantitatively assess the performance of our res-cycle GAN based INU correction algorithm, two profiles were plotted as shown in the bottom row in Fig. 3. In the bottom-left of Fig. 3, the profiles were plotted along the red dashed line in the top row images, where severe INU exists in the uncorrected MRI image. From the profiles, we can see a large intensity ripple (blue in the bottom-left in Fig. 3) within a similar type of tissues. This intensity non-uniformity was corrected very well by either our proposed method (red) or N4ITK (green). On the other hand, profiles were plotted across the regions with little INU (as shown as the blue dashed lines in the images in the top row in Fig. 3). From the curves in the bottom-right of Fig. 3, both our proposed method and N4ITK can preserve the intensity flatness within a similar type of tissue as original raw MRI. The res-cycle GAN based method can achieve excellent INU correction performance at the same time avoiding over correction as the N4ITK algorithm.
	Fig.4 shows axial (top row), sagittal (second row), and coronal (third row) cross-sectional images of uncorrected, N4ITK corrected, and correction by our proposed method, respectively. The differences between N4ITK and the proposed method were shown in the last column in Fig. 4. From these images, we can see, similar INU correction performance as N4ITK was achieved by the proposed res-cycle GAN based method.
	\begin{figure}
		\centering		
		\noindent \includegraphics*[width=6.50in, height=4.20in, keepaspectratio=true]{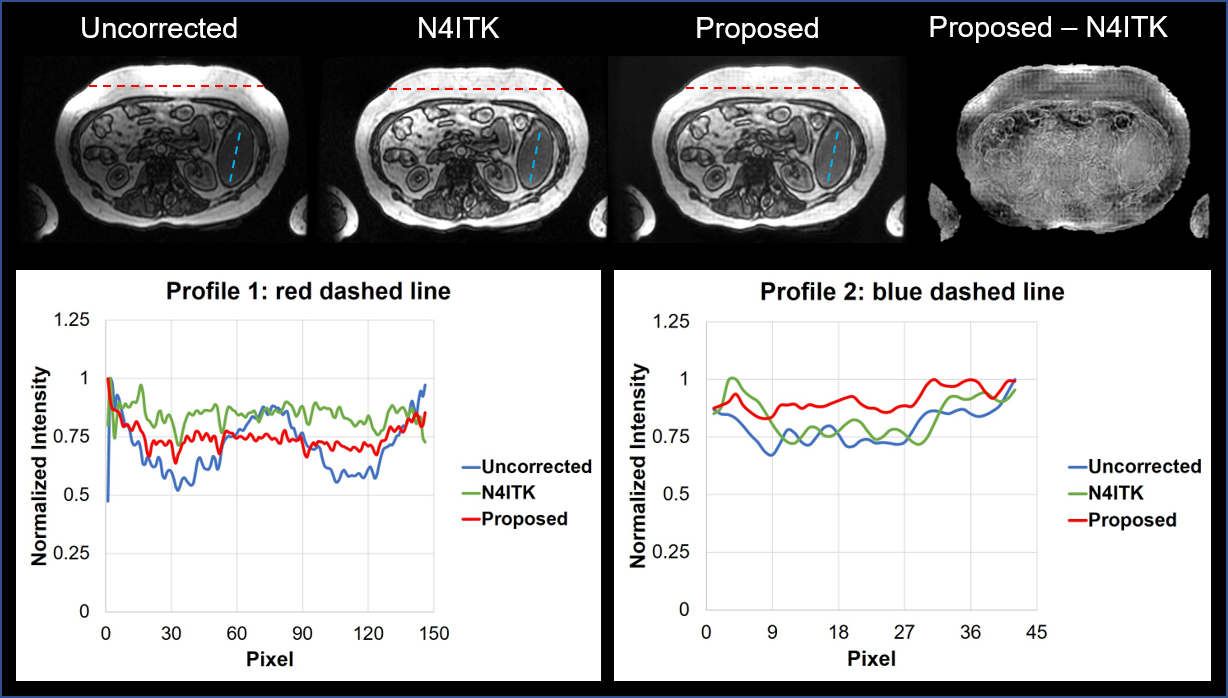}
		
		\noindent Fig. 3. Summary of INU correction results in one patient.
	\end{figure}

	\begin{figure}
		\centering		
		\noindent \includegraphics*[width=6.50in, height=4.20in, keepaspectratio=true]{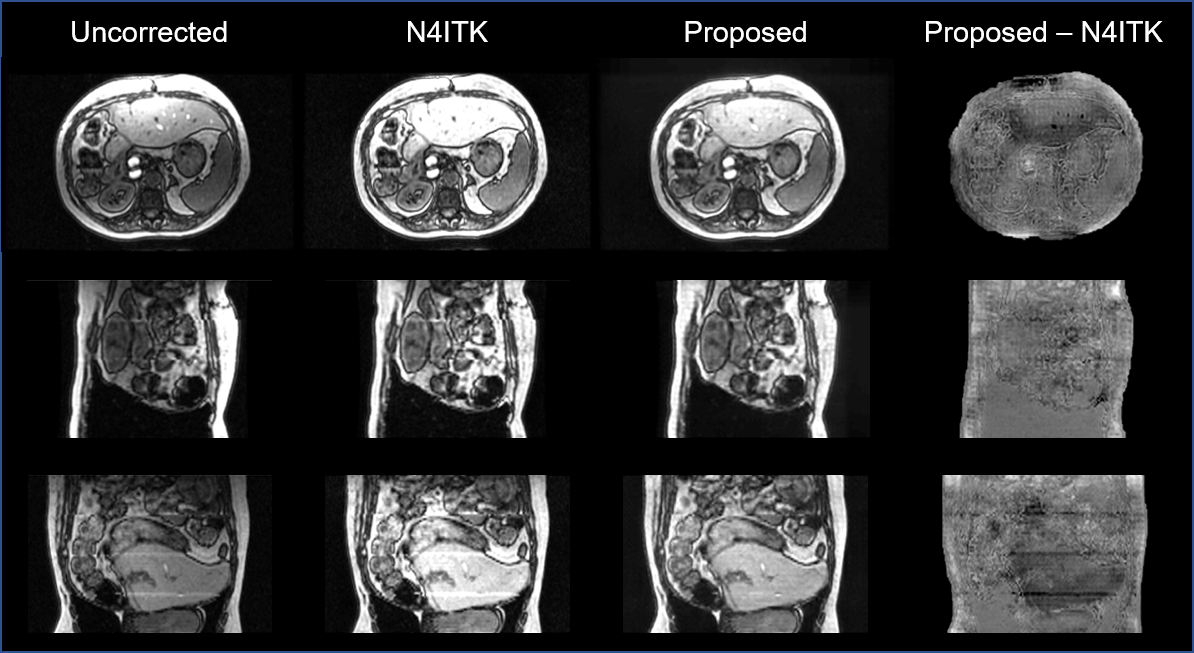}
		
		\noindent Fig. 4. Three orthogonal views of the uncorrected, N4ITK corrected, res-cycle GAN based corrected, and the differential between proposed and N4ITK corrected MR images.
	\end{figure}
	
	\subsection{Comparison to other machine learning-based methods}
	
	The proposed res-cycle GAN based correction method is compared to two other popular networks (GAN\cite{RN3824}, and U-net\cite{RN3811}) based correction methods. As shown in Fig. 5, the first row shows the images of uncorrected, N4ITK corrected, proposed res-cycle GAN based corrected, GAN based corrected, and U-net based corrected, respectively. The second row of Fig. 5 shows the estimated non-uniformity fields calculated by N4ITK, res-cycle GAN, GAN, and U-net, respectively. To quantitatively compare among these methods, the differential images were obtained between the output images from res-cycle GAN, GAN, and U-net, and their ground truth during the network training process, which is N4ITK corrected images. As shown in the third row in Fig. 5, the performance of our proposed method is better than either GAN or U-net based INU correction method. 
	The three top rows in Fig. 6 shows three orthogonal views of the uncorrected, N4ITK corrected, and res-cycle GAN based corrected, GAN based corrected, U-net based corrected MR images of one patient. Again, their differential images with their ground truth images (N4ITK corrected) were calculated and shown in the fourth to sixth rows in Fig. 6. Similar better performance of our proposed method over GAN or U-net based correction method is seen from three orthogonal views (axial, sagittal, and coronal).

	\begin{figure}
		\centering
		\noindent \includegraphics*[width=6.50in, height=4.20in, keepaspectratio=true]{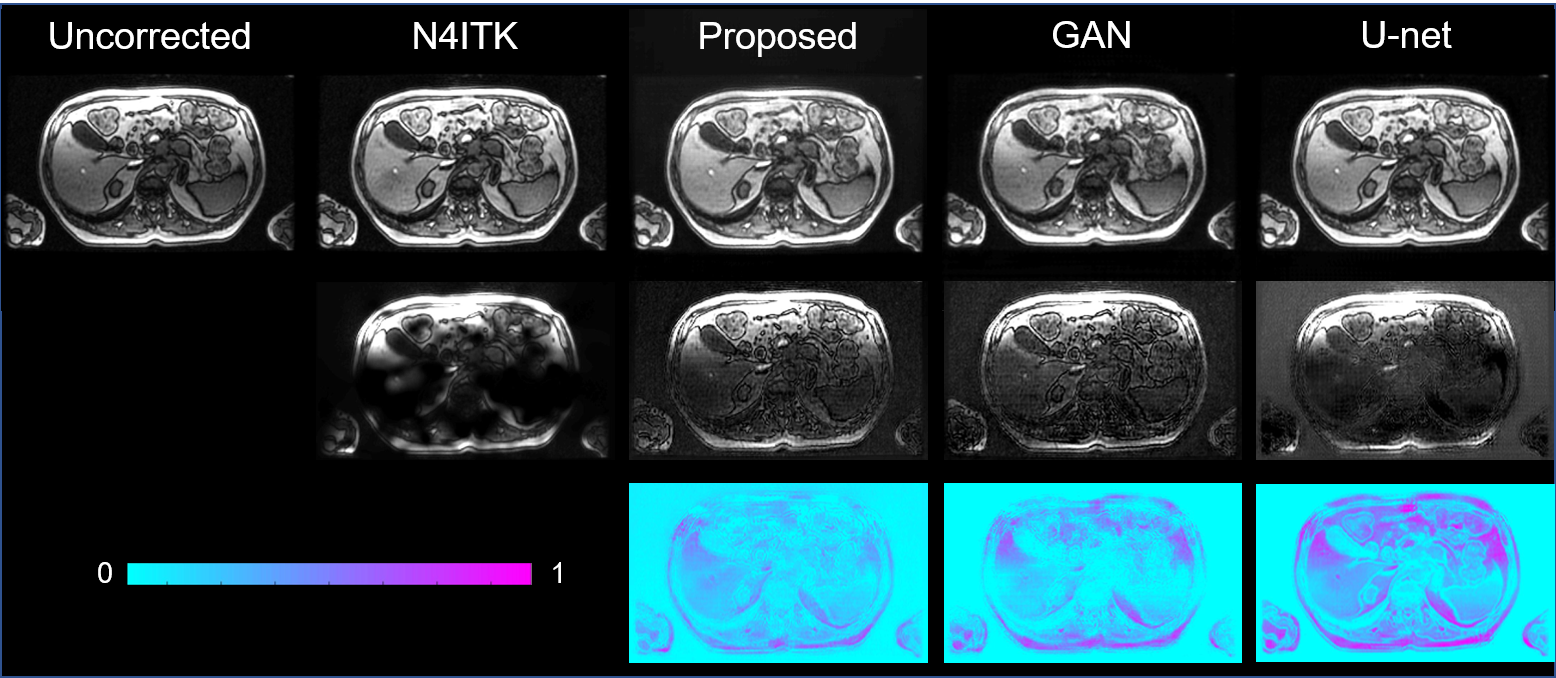}
		
		\noindent Fig. 5. Summary of INU correction results by N4ITK, res-cycle GAN, GAN, and U-net.
	\end{figure}

	\begin{figure}
		\centering
		\noindent \includegraphics*[width=6.50in, height=4.20in, keepaspectratio=true]{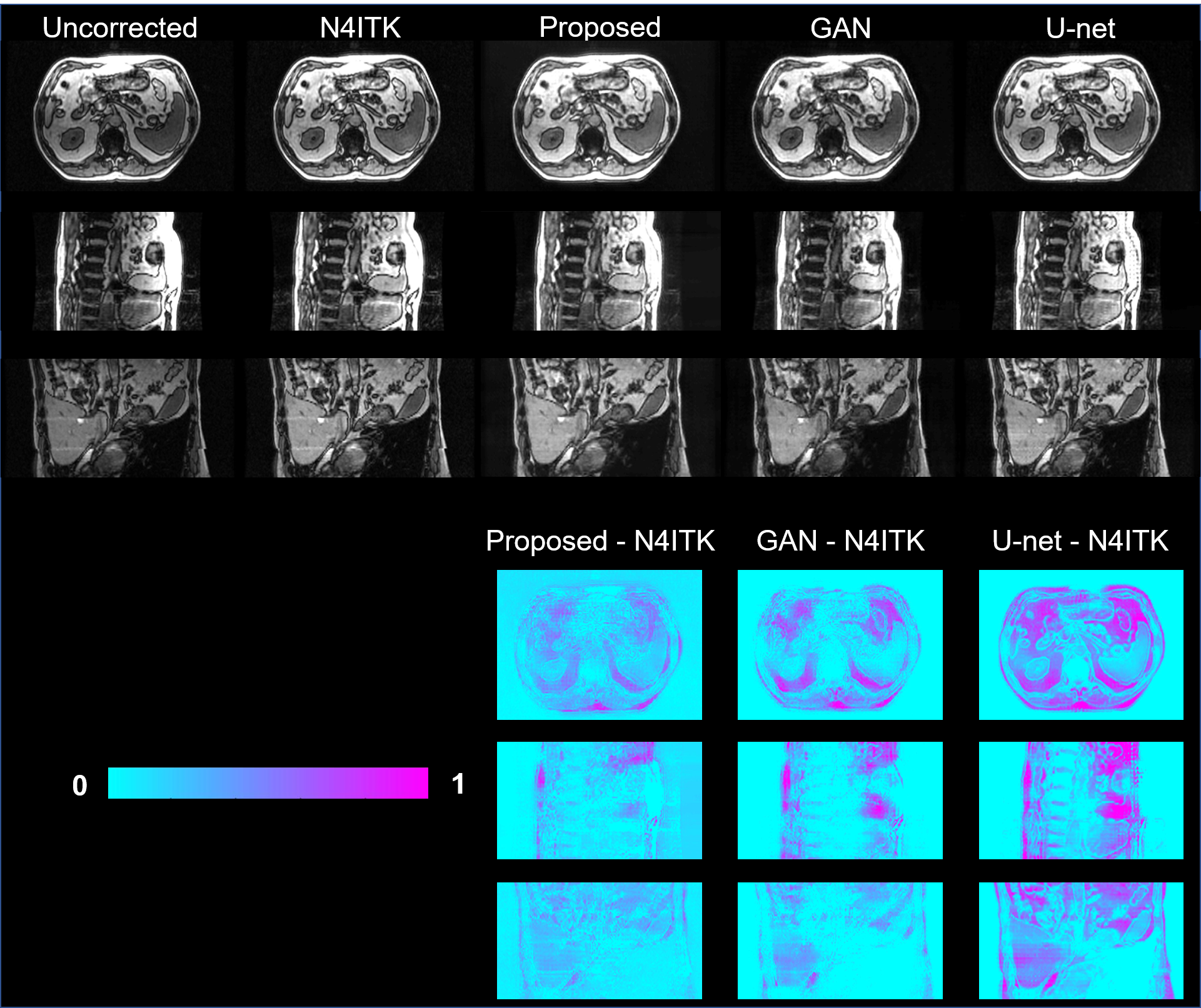}
		
		\noindent Fig. 6. Three orthogonal views of the uncorrected, N4ITK corrected, and res-cycle GAN based corrected, GAN based corrected, U-net based corrected MR images.
	\end{figure}
	
	\subsection{Quantitative comparison among different INU correction methods}
	The quantitative results of NMAE, PSNR, NCC, SNU among uncorrected, our res-cycle GAN, GAN, and U-net for all test cases are summarized in Table 1. As is shown in Table 1, our proposed method outperformed GAN, and U-net based methods in NMAE, PSNR, NCC and SNU. While all methods improved upon the original uncorrected image, the proposed res-cycle GAN based was found to be superior to either GAN or U-net based INU correction method.

\begin{table}[htbp]
	\centering
	\caption{Quantitative results and p-value obtained by comparison among the proposed method, U-net based, and GAN based method. For all tests, the N4ITK corrected image was taken as the ground truth.}
	\begin{tabular}{cccccc}
		\toprule
		\toprule
		& NMAE& PSNR(dB) & NCC & SNU\\
		\midrule
		Uncorrected & 0.032±0.019 & 25.9±5.6 & 0.930±0.035 & 0.747±0.173 \\
		U-net & 0.018±0.007 & 24.1±2.7 & 0.964±0.029 & 0.562±0.156 \\
		GAN & 0.012±0.002 & 26.7±1.3 & 0.956±0.037 & 0.424±0.093 \\
		Proposed & 0.011±0.002 & 28.0±1.9 & 0.970±0.017 & 0.298±0.085 \\
		\makecell{P-value\\(U-net vs. proposed)} & $<$0.001 & $<$0.001 & 0.200 & $<$0.001 \\
		\makecell{P-value\\(GAN vs. proposed)} & 0.016 & $<$0.001 & 0.012 & $<$0.001 \\
		\bottomrule
		\bottomrule
	\end{tabular}%
\end{table}%

	\bigbreak
	
	\noindent 
	\section{Discussion}
	
	Artifacts reduction is an important preprocessing step in most MR image analysis such as image segmentation, image registration, feature extraction, radiomics, and so forth. The proposed res-cycle GAN based method is well suited for intensity non-uniformity correction in MRI. The results show that the proposed method achieves better performance than conventional learning-based methods including two popular networks, GAN and U-net based methods. In original cycle-consistent GAN, unpaired images were investigated, while the proposed algorithm used paired training data (paired uncorrected and N4ITK corrected images). By using paired data, in which, the differences between each pair of images are primarily the INU artifacts, thus, it allows the algorithm to primarily focus on reducing image INU artifacts, rather than focusing on other distortions such as motion artifacts, geometric distortions, and so forth. To some extent, it improves the INU correction performance of our proposed method. Moreover, the residual learning strategy was also adopted and integrated to the cycle-consistent GAN, which speeds up the training process and enhances the convergence of training. From this point of view, the proposed res-cycle GAN method has more efficiency than the original cycle-consistent GAN, conventional GAN, and U-net based methods. 
	
	A limitation of the study is the ground truth used for training the network was N4ITK correction. N4ITK is an established and widely used INU method in MRI. It can provide images with highly INU reduction compared to uncorrected images in MRI. However, in practice, INU is a complex issue caused by many non-ideal aspects of MRI including static field inhomogeneity, radio frequency coil non-uniformity, gradient-driven eddy currents, inhomogeneous reception sensitivity profile, and overall subject’s anatomy both inside and outside the field of view, and so forth, which cannot be fully corrected by N4ITK. And the parameter tuning process in N4ITK correction method is necessary and will affect the correction results. Therefore, there’re unavoidable errors in N4ITK corrected images compared to ideal INU free images. It would be better to use ideal INU free images as ground truth to train the network in our proposed res-cycle GAN based method. However, in practice, it is difficult to obtain ideal INU free images. Thus, without loss of generality, in this study, N4ITK corrected images were still recognized as ground truth for training networks. The results show that the proposed method can provide INU corrected images as close as the ground truth (N4ITK corrected). It demonstrates that the proposed method can achieve excellent correction performance. 
	
	As technical advances in artificial intelligence (AI), more and more AI-based methods will have higher and higher impacts on traditional diagnostic imaging and therapeutics. Highly automated medical image processing and analysis methods will greatly release human labor in a daily basis. This study offers a generic framework for image preprocessor or artifacts reduction in medical image processing and analysis, especially, in the aspects of removing artifacts caused by intensity non-uniformity. Compared to other traditional methods like N4ITK, the proposed method can be easily customized and integrated to the image analysis methods such as image segmentation algorithms, image registration methods, and so forth.

	\bigbreak
	
	\noindent 
	\section{Conclusion}
	
	In this study, an advanced deep learning method, namely, 3D residual-cycle-GAN were implemented and validated for intensity non-uniformity correction in MRI. With the proposed method, highly automated intensity non-uniformity correction in MRI is achievable. It avoids time-expensive unintuitive parameter tuning process in N4ITK correction method, which is currently the most well-known, efficient, and the most commonly used method for intensity non-uniformity correction in MRI. Moreover, our proposed method is capable of capturing multi-slice spatial information which results in a smoother and uniformed intensity distribution in the same organ compared to the N4ITK correction. Besides, the proposed method outperforms other learning-based methods including GAN, and U-net based INU correction methods. Quantitative comparisons including intensity profile plots, normalized mean absolute error (NMAE), peak signal-to-noise ratio (PSNR), normalized cross-correlation (NCC) indices, and spatial non-uniformity (SNU) were made among the proposed method and other approaches. The results show that our proposed method can achieve higher accuracy than other methods. Compared to N4ITK correction, our proposed method highly speeds up the correction through avoiding the unintuitive parameter tuning process in N4ITK correction. Moreover, the proposed method further removes the artifacts in the preprocessing step, which will help improve the performance of quantitative image analyses including segmentation, registration, classification, and feature extraction, thus provide more accurate clinical essential information for diagnosis and therapy.

	\noindent 
	\bigbreak
	{\bf ACKNOWLEDGEMENT}
	
	This research is supported in part by the National Cancer Institute of the National Institutes of Health under Award Number R01CA215718 and R01EB028324, and Dunwoody Golf Club Prostate Cancer Research Award, a philanthropic award provided by the Winship Cancer Institute of Emory University.

	\noindent 
	\bigbreak
	{\bf Disclosures}
	
	The authors declare no conflicts of interest.

	\noindent 
	
	\bibliographystyle{plainnat}  
	\bibliography{arxiv}      
	
\end{document}